\documentclass[twocolumn, prb, showpacs, superscriptaddress]{revtex4-1}
\usepackage{amsmath,amssymb,epsfig,color}
\usepackage{graphicx}
\usepackage{dcolumn}
\usepackage{bm}
\usepackage{color}

\begin{document}

\title{Raman evidence for nonadiabatic effects in optical phonon self-energies\\ of transition metals}

\pacs{74.25.nd, 74.70.Ad, 63.20.dk, 63.20.kd }

\author{Yu. S. Ponosov}
\affiliation{M. N. Mikheev Institute of Metal Physics UB RAS, S. Kovalevskaya
Str. 18, 620990  Ekaterinburg, Russia} \affiliation{Ural Federal
University, Mira Str. 19, 620002 Ekaterinburg, Russia}
\author{S. V. Streltsov}
\affiliation{M. N. Mikheev Institute of Metal Physics UB RAS, S. Kovalevskaya
Str. 18, 620990 Ekaterinburg, Russia} \affiliation{Ural Federal
University, Mira Str. 19, 620002 Ekaterinburg, Russia}

\begin{abstract}

We report a Raman study of the effect of temperature on the self-energies of
optical phonons in a number of transition metals with hexagonal-close-packed
structure. Anisotropic softening of phonon energies and narrowing of phonon
linewidths with increasing temperature are observed. These effects are
reproduced in the calculations of phonon spectral functions based on
\textit{ab initio} electronic structures and with carrier scattering by
phonons taken into account. The combined observations and results of
simulations indicate a relation between observed anomalies and the
renormalization of the electron spectrum due to electron-phonon interaction.
It is emphasized that the temperature dependence of the phonon energies
resembles anharmonic behavior but is actually an electron-induced effect.

\end{abstract}

\date{\today}

\maketitle

\section{Introduction}

The adiabatic approximation is commonly applied in the description of the
phonon spectra of metals. Nonadiabatic (NA) corrections due to
electron-phonon interaction usually are small, of the order $\sqrt{m/M}$
where $m$ and $M$ are the electronic and ionic mass, respectively~\cite{mig}.
However, Engelsberg and Schrieffer~\cite{eng} were first to show that the
nonadiabatic renormalization of adiabatic frequencies of long wavelength
optical phonons can be larger than $\sqrt{m/M}$ if the phase velocity of
phonons becomes larger than the electronic Fermi velocity $\upsilon_f$,
\begin{equation}\label{eq1}
\frac{\omega_0}{\textit{q}\upsilon_f}\geq 1.
\end{equation}
Here $q$ is the phonon wavevector and $\omega_0$ is the phonon frequency.
Correspondingly,  the electron-phonon
interaction should be much enhanced for low-$q$ phonons from the NA regime. This
may increase the superconducting transition temperature $T_{c}$ in correlated
electron systems~\cite{kul}. In the past years it has been
proposed~\cite{cap, alex,cal} that nonadiabatic effects in electron-phonon
interaction could be responsible for the high-temperature superconductivity
in compounds with low carrier density such as fullerenes~\cite{ful} or
Mg$B_{2}$~\cite{MgB}. The recent measurement of Raman spectra in graphite
intercalation compounds ~\cite{CaC, intG} showed that the phonon frequencies
related to in-plane carbon vibrations are significantly larger than those
obtained from density functional theory (DFT) calculations in the adiabatic
approximation. These findings have been explained by giant NA effects (up to
30$\%$ in phonon energy) calculated for a number of intercalated graphite
systems and 3D metals~\cite{mauri} where the interband mechanism  was
proposed to be the main source of nonadiabaticity.

In metals, nonadiabatic effects related to the electron-phonon interaction
were shown to result in a singularity in the optical phonon branch for $\textit{q} \upsilon_f \approx \omega_0$ which
depends on the coupling constant~\cite{ip,maks}. At $q_0$ =
$\omega_0/\upsilon_f$ the threshold for Landau damping for an optical phonon
has been predicted and the temperature behavior of the phonon frequencies and
linewidths has been discussed~\cite{maks,ip2,maks1}.

In contrast to neutron inelastic scattering, Raman spectroscopy only covers
the near-zone-center regime ($10^5-10^6$~cm$^{-1}$) but offers high
resolution. The experimental detection of NA effects in metals, however, is
rather difficult because of smearing of the probed wavevector. It
results from the finite penetration depth of the laser light and leads to the
observation of both adiabatic and NA contributions to optical phonon
frequencies and linewidths. Such effects of spatial dispersion on the
electron-phonon interaction were first found in pure hexagonal-close-packed (hcp) metals with long mean
free path~\cite{pon1}. There, the coupling is indicated by an anomalous
temperature behavior of the $E_{2g}$ phonon linewidths. Along with this anomaly,
the phonon frequencies showed a typical softening upon
cooling~\cite{Re,Os,Ti}.

In this report we present the results of new experimental studies of  NA
effects  in a number of hcp transition metals. We
confirm that in these anisotropic metals the NA effects lead to a variation of
optical phonon energy being dependent both on the value and direction of the
phonon momenta~\cite{pbtc}. We simulated the spatial dispersion
effects and the anomalous temperature behavior of the phonon linewidth  in
our model calculations.  They are based on density functional theory (DFT) electronic
structures and take into account intraband transitions. It is shown
that renormalization of the bare electron spectrum due to electron-phonon
interaction is the main mechanism of phonon self-energy changes. In
particular, it is specified that the phonon softening with increasing temperature
is not the usual anharmonic effect but is due to the electron-phonon
interaction.

\section{Experiments}

For temperature-dependent measurements, electro-polished plates of Y, Zr, Ti,
Ru, and Os single crystals (residual resistance ratio $\geq 50$) with
different orientations of the scattering surface were placed into an optical
cryostat (Oxford or Cryovac).  Most measurements were performed with
excitation by laser radiation of wavelengths 514~nm and 633~nm. The focal
spot on the sample varied from 2 to 10~$\mu$m diameter, depending on
settings. For Ru and Os, additional lines of a Kr ion laser were used. Raman spectra were
recorded by Renishaw and Labram microscope spectrometers with the spectral
resolution (approximated by a Gaussian shape) of about 3~cm$^{-1}$. In order
to obtain the  intrinsic phonon frequencies and linewidths, the measured
spectra were fitted with a Voigt  expression.

\section{Calculations}

The Raman line shape in metals is affected by both anharmonicity and
electron-phonon coupling. Calculations of the phonon spectral function were
carried out in order to theoretically estimate the effects of electron-phonon
interaction on the phonon energies and linewidths of the investigated metals. As
a result of the wavevector smearing of the light field in a metal and in the
presence of strong dispersion of the frequency and damping, a measured phonon
line shape may have a rather complicated form. The phonon spectral function
$I(\omega)$ was calculated by taking into account the frequency and
temperature dependences of the phonon self-energies $\Pi(q,\omega,\omega
_0)$ :
\begin{equation}\label{eq2}\begin{split}
&I\left(\omega\right) = \frac{4\omega_0^{2}}{\pi}\; \times\\
&\int_{0}^{\infty} dq \frac{{U(q) \Gamma \left( {q,\omega ,\omega _0 }
\right)}}{{\left[ {\omega ^{2}  - \omega
_{0}^{2}  - 2\omega_0 \Pi^{\prime}\left(q,\omega,\omega _0\right)} \right]^{2} +
4\omega_0^2 \Gamma^2\left(q,\omega ,\omega _0\right)}}
\end{split}\end{equation}
Here $\omega_0$ is the bare (uncoupled) phonon energy,
$\Gamma(\vec{q},\omega,\omega_0) = \Gamma_0 + \Pi''(\vec{q},\omega)$ is the
total width, $\Pi'(\vec{q},\omega)$ and $\Pi''(\vec{q},\omega)$ are the real
and imaginary parts of the phonon self-energy. The bare line width $\Gamma_0$
includes zero-temperature contributions from defects, anharmonicity, etc. The breakdown of optical wavevector conservation in a metal leads to coupling of the light to phonons with a spread of wavevectors and
hence of frequencies. Following~\cite{lou} we take $U$(q) to have the form
$U(q)\propto4q^{2}/\left|q^{2}-\xi^{2}\right|^{2}$, where 
$\xi=\xi_{1}-i\xi_{2}=(2\omega_{i}/c)\times(n-ik)$.  For incident laser energy $\omega_{i}$ probed wavevector distribution $U$(q) respresents a skew lineshape with a peak position $q_{i}=(2\omega_{i}/c)\times\sqrt{n^{2}+k^{2}}$ and was estimated on the basis of experimental refraction index n and extinction coefficient k for investigated metals~\cite{kny,kir}. 

The phonon self-energy originates from intraband
electronic transitions in our simulation. We used $\Pi(q,\omega)$ in the form:~\cite{maks}
\begin{equation}\label{eq3}\begin{split}
&\Pi \left( {q,\omega ,\omega _0 } \right) = \oint
\frac{ds_f}{\upsilon_f } \; g^{\rm{2}} \left( {k_f,
q, \omega_0 } \right)\; \times\\
& \left\{ \left(  \int\limits_{ - \infty }^\infty  {d\varepsilon }
 \frac{f\left( \varepsilon  \right) - f\left(\varepsilon  + \omega  \right)}{\omega  - q \upsilon_f^z  - \Sigma^A
\left( \varepsilon  \right) + \Sigma^R\left( \varepsilon  + \omega \right)}\right)- {\rm{2}} \right\}.
\end{split}\end{equation}
Here, $ds_f$ is the area element of the Fermi surface, $\upsilon_f$ - the
electron velocity, g - the matrix element of the electron-phonon interaction,
$f(\varepsilon)$ is the Fermi function, and $z$ denotes the normal to the
sample surface. The advanced and retarded quasi-particle electronic
self-energies $\Sigma^A(\epsilon)$ and $\Sigma^R(\epsilon+\omega)$ determine the
electron spectrum renormalization near the Fermi level due to different
interactions. In the case of electron-phonon scattering, the real and
imaginary parts of $\Sigma(\epsilon)$ are~\cite{su}
\begin{equation}\label{eq5}\begin{split}
&\Sigma^{'}(\epsilon)= \int d\Omega\alpha^2F(\Omega) \; \times\\
&\Re \left[ \psi \left( \frac{1}{2}+i\frac{\epsilon+\Omega}{2T}\right)-\psi\left(\frac{1}{2}+i\frac{\epsilon-\Omega}{2T} \right) \right],
\end{split}\end{equation}
\begin{equation}\label{eq6}\begin{split}
&\Sigma^{''}(\epsilon)= \pi \int d\Omega\alpha^2F(\Omega) \; \times\\
&\left[ 2n_B(\Omega)-f\left(\epsilon-\Omega\right)+f\left(\epsilon+\Omega\right)+1 \right]+\nu.
\end{split}\end{equation}
Here $\nu$ is the electron-impurity scattering rate, $\Psi$ the digamma
function, $\Omega$ the phonon energy, $\alpha^2F(\Omega)$ the Eliashberg
spectral function for the electron-phonon interaction, and $n_{\rm B}$ the
Bose function. The phonon densities of states $F(\Omega)$ from
Refs.~\cite{st,Ru} or from measured second order Raman spectra (in case of Os, see Fig.S1) were used in
our calculations.

The electron velocities on the Fermi surface have been obtained from
band-structure calculations that use the linearized muffin-tin orbital method
(TB-LMTO-ASA)~\cite{and}) in the local density approximation. The obtained
Fermi surfaces are in agreement with those calculated earlier~\cite{jep}.
Integration over the Fermi surface was performed with a fine mesh of 125\,000
k-points in the full Brillouin zone.

\section{Results}

\subsection{Group III metal yttrium}
\begin{figure}[tb] \includegraphics[width=80mm]{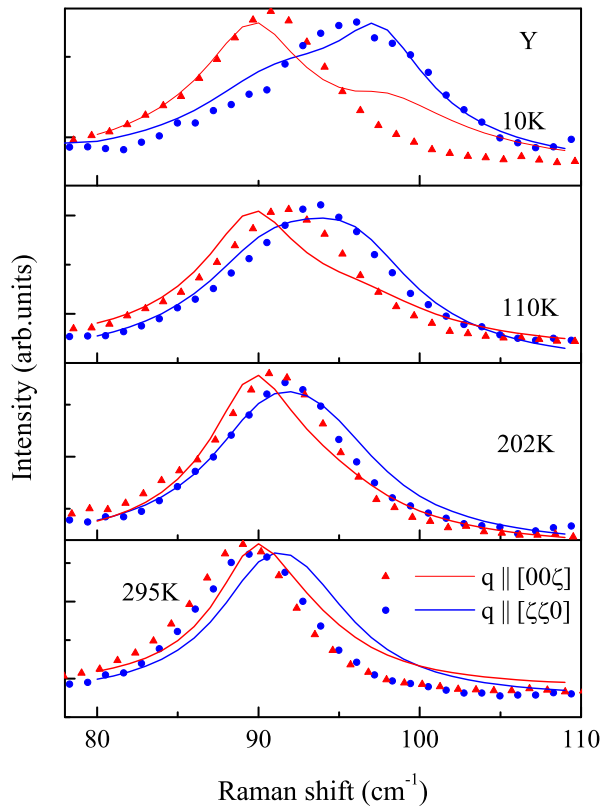}
\caption{\label{fig1} Measured (symbols) and calculated (lines) phonon spectral functions of yttrium at different temperatures for two directions of the wavevector $q$. The excitation wavelength was 633 nm.} 
\end{figure}

\begin{figure}[b]
\includegraphics[width=80mm]{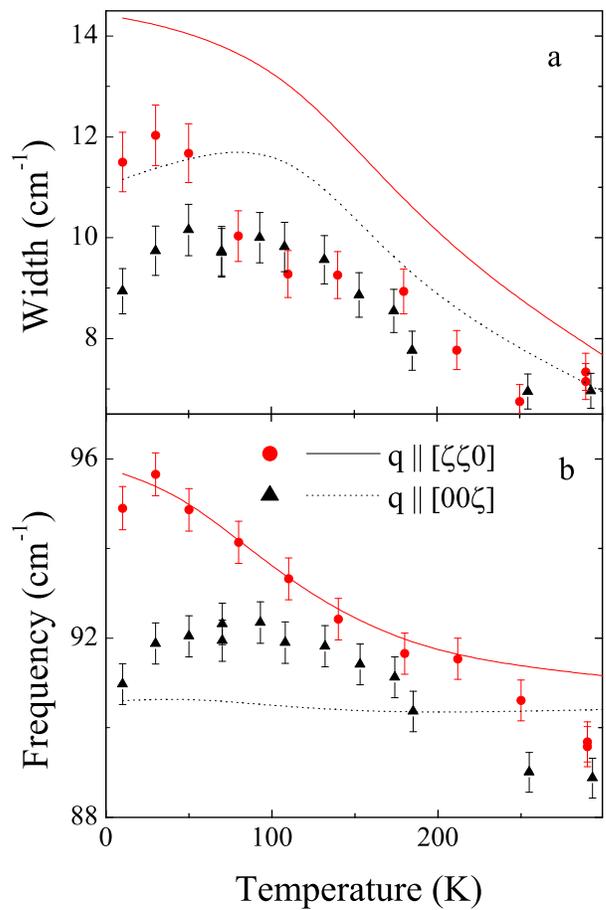}
\caption{\label{fig2}Experimental (symbols) and calculated (lines) temperature dependence of the phonon linewidth (a) and frequency (b) in yttrium for two wavevector directions. The excitation wavelength was 633 nm.}
\end{figure}

Figure~\ref{fig1} shows measured and calculated lineshapes and Fig.~\ref{fig2} present the energy and width of the
$E_{2g}$ optical mode in yttrium as a function of temperature. The phonon frequencies for probed
\textit{q} directions, cf. Fig.~\ref{fig1} and Fig.~\ref{fig2}, exhibit a $\simeq$ 5~\% difference at low
temperatures. The difference almost disappears at room temperature. Low-temperature
linewidths, cf. Fig.~\ref{fig1} and Fig.~\ref{fig2}, are also different and show non-monotonic temperature behavior
with maximum near 100~K for q$\|$[00$\xi$], then decreasing with further 
increase of temperature. There are no low-temperature phase transitions in
Y. It is therefore natural to assume that the observed anisotropy of the phonon frequencies and widths and unusual temperature dependence of the latter is due to the electron-phonon interaction.

Using Eqs.~\ref{eq2} -- \ref{eq6} we performed calculations
of phonon line shapes for different wavevector directions and temperatures.
In these and the following calculations we used values of $\Gamma_0
=0.5-0.7~{cm}^{-1}$ and $\nu = 10~{cm}^{-1}$. Adjusted parameters  were
$\omega_0$,  $g$ and an electron-phonon  coupling constant $\lambda = 2\,\int d \Omega
\alpha^2F(\Omega)/\Omega$. Because  the phonon energies in all the studied
metals do not depend on the $q$ direction at room temperature, this allows to
vary the bare frequency  $\omega_0$ and the matrix element of electron-phonon
interaction $g$ in narrow limits.  Using literature values of
$\lambda_{tr}$=0.62~\cite{al} or calculated $\lambda=0.55$~\cite{singh}
couldn't provide any acceptable agreement with experimental results. The
best agreement was obtained using  $\lambda=0.13$.  Parameters used in the calculations for the
five metals are listed in Table~\ref{tab1}.
\begin{table}[tb]
\caption{\label{tab1} Parameters used in calculations of phonon self-energies
of the investigated metals.}
\begin{ruledtabular}
\begin{tabular}{llllll}
  & $\omega_0$ (cm$^{-1}$) & $g$ (cm$^{-1}$) &$\Gamma_0$ (cm$^{-1}$) &$\lambda$ &$\lambda_{E_g}$\\
Y & 97&95&0.5&0.13&0.056\\
Zr & 104&145&0.5&0.49&0.059\\
Ti & 157&203&0.7&0.54&0.064\\
Ru & 213&200&0.5&0.66&0.042\\
Os & 179&213&0.5&0.72&0.042\\
\end{tabular}
\end{ruledtabular}
\end {table}

\begin{figure}[t] \includegraphics[width=90mm]{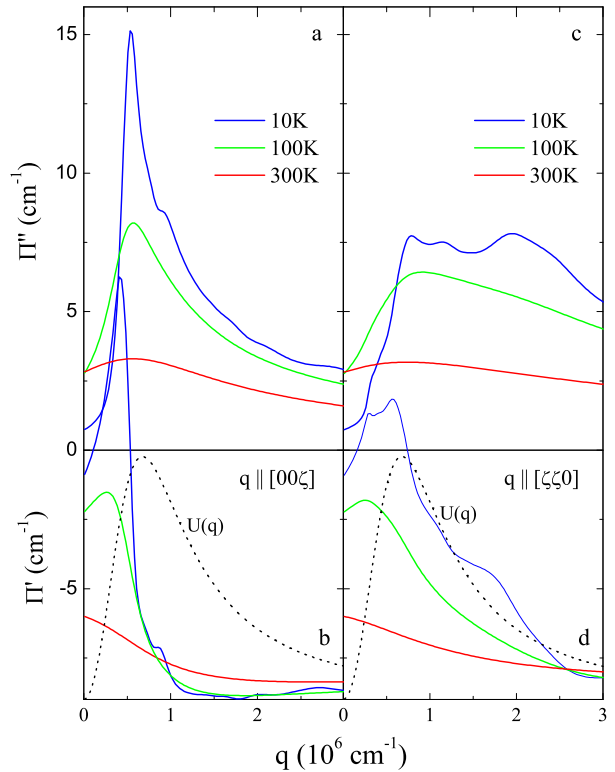}
\caption{\label{fig3}  Calculated imaginary (a,c) and real (b,d) parts of the phonon self-energy of Y versus wavevector $q$ for two q-directions and different temperatures. Dotted line in (b) and (d) shows probed wavevector distribution for excitation wavelength 633 nm. } 
\end{figure}

The $q$-dependences of the calculated phonon self-energies are shown in
Fig.~\ref{fig3} together with probed wavevector distribution. It follows from Fig.~\ref{fig3} that the calculated $q_0$ value for the Landau damping threshold and strong dispersion region for q $\|$ [00$\xi$] is
approximately two times less than for the in-plane $q$ direction. The value of $q_0$ was evaluated at bare phonon frequency $\omega_0$ (Table~\ref{tab1}) from Eq.3  where we used isotropic electronic self-energies $\Sigma$. Therefore, the low-temperature difference of $q_0$ and phonon energies for different q-directions is due to the  difference of average bare velocities $\upsilon_{f}^{z}$. In its turn, the latter are determined by both a bare electron velocity and a shape of the Fermi surface, which sets the value of the average velocity for a certain q - direction. The result shown for yttrium in Fig.~\ref{fig3} is commented in Supplemental information (see Fig.S2).

Therefore, having in mind the wavevector distribution, it becomes clear that
for q $\|$ [00$\xi$]  the adiabatic region provides the main contribution to the
phonon frequency. For the in-plane $q$ direction the contribution of the NA region prevails,
where the phonon frequencies are higher at low temperatures. Strong
renormalization of the electron mass (velocity) in the NA region~\cite{maks} leads to the
phonon frequency decrease when temperature increases. Although such temperature
trend is qualitatively similar to anharmonic behavior, the actual reason for phonon softening is a disappearance
of NA contribution at high temperatures. The anomalous narrowing of the
phonon linewidth with increasing temperature can be attributed to a decrease of
the number of electron states which strongly interact with phonon and provide its damping both in adiabatic and nonadiabatic q-regions.  The electron lifetime in the NA region decreases due to an electron-phonon scattering which leads to the increase of the
phonon width  at intermediate temperatures~\cite{ip2,maks,mauri}. The interplay between these opposite trends is
manifested sometimes in the non-monotonic dependences of the phonon widths.

The solid lines in Fig.~\ref{fig2}(a,b) show the frequencies and widths
obtained from fitting of the phonon spectral functions calculated using
phonon self-energies of Fig.~\ref{fig3}. The agreement
with experimentally observed tendencies is rather satisfactory though our
simple calculation uses constant and isotropic $\alpha^2(\Omega)$ and $g$.  It should be noted
that phonon spectral functions calculated for low temperatures have rather
complicated shapes because of wavevector smearing. They show shoulders originating from
 NA region of the spectrum (for q $\|$ [00$\xi$]), when the main contribution is determined by 
the adiabatic region or, conversely, for the in-plane $q$ direction, when the peak from NA 
region dominates (Fig.~\ref{fig1}). Experiment actually finds asymmetric line shapes but the presence of the second
peak is less pronounced in most cases (Fig.~\ref{fig1}). As a rule, phonon
peaks become symmetric at high temperatures. We have not tried to decompose
the experimental and calculated peaks for the components because the
resulting form is a complex integral function of the probed distribution of
the wavevectors  and anomalous phonon dispersion. Instead, both the
experimental and calculated line shapes were fitted by a Voigt (Lorentz)
distribution to determine the central frequency and linewidth. Unfortunately, the use of such 
a procedure for fitting the calculated spectral functions sometimes leads to frequency shifts 
and band broadening due to the appearance of double peaks. 

It should be noted that little information on the $\lambda$ magnitude of Y
exists. Our estimate, indicating weak electron-phonon interaction in spite of
high density of states at the Fermi level, is in agreement with absence of
superconductivity in Y at atmospheric pressure. Estimated mode constant
$\lambda_{E_{2g}}$=$g^2N_F/\omega_0\simeq$ 0.056, where $N_F$ is a total
density of states at the Fermi level.

\subsection{Group IV metals -zirconium and titanium}
\begin{figure}[t] \includegraphics[width=80mm]{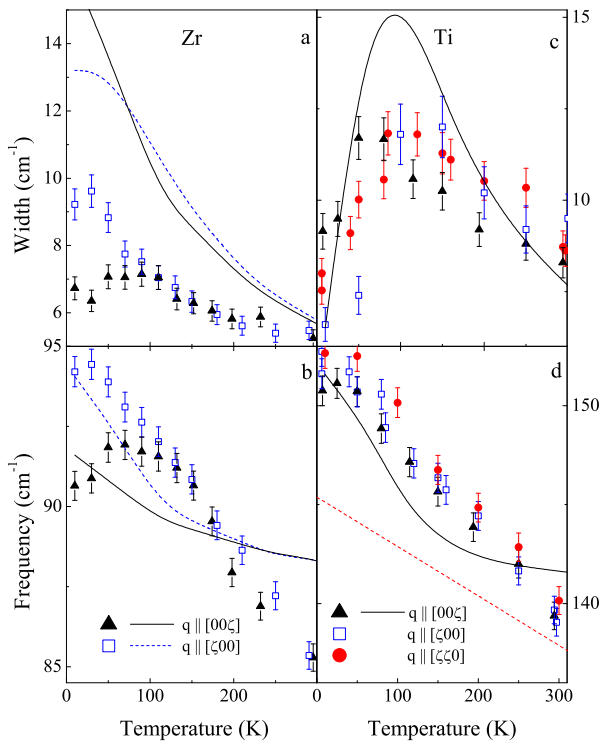}
\caption{\label{fig4}  Experimental (symbols) and calculated (lines)
temperature dependence of the phonon linewidth and frequency in zirconium (a,b)
and titanium (c,d) for different directions of the wavevector $q$. The excitation wavelength was 633 nm. Dashed line in (d) is interpolation of high-temperature neutron data~\cite{st1} for $E_{2g}$ mode in Ti. } 
\end{figure}

The measured phonon lineshapes in the 
group IV metals Zr and Ti are shown in Fig.S3. The temperature
dependences of the $E_{2g}$ frequencies and widths in
 Zr and Ti are presented in Fig.~\ref{fig4}. The results for Zr
are similar to those for Y and show differences in
frequencies and widths for in-plane and \textit{c}-axis $q$ directions at temperatures $\triangleleft$ 100K (Fig.4).
 Our calculations reproduce the tendency of
anisotropic phonon frequency softening and width narrowing with increasing
temperature. The calculated linewidth for momentum along an \textit{c} axis at low temperatures exceeds the linewidth for  the momentum in the basal plane in contrast to the experiment (Fig.~\ref{fig4}(a,b)). The reason is an
appearance of a shoulder on the high frequency side of the phonon peak. This leads to upward shift of the frequency and peak broadening with accepted procedure for fitting of calculated spectral function.

\begin{figure}[t] \includegraphics[width=80mm]{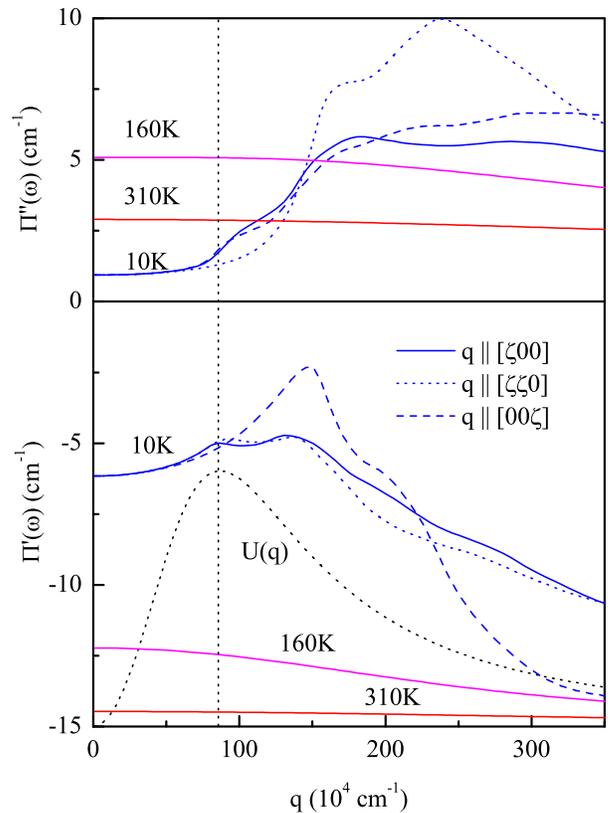}
\caption{\label{fig5} Calculated  real (a) and  imaginary (b) parts of the phonon self-energy of Ti versus wavevector $q$ for different  q-directions and temperatures. Dotted line in (b) and (d) shows probed wavevector distribution for excitation wavelength 633 nm.}
\end{figure}

Another situation is observed in Ti where the phonon parameters were found to
be isotropic for all three $q$ directions. In Fig.~\ref{fig4}(c,d) we present results for q $\|$
[00$\xi$] and q $\|$ [$\xi\xi$0] in addition to earlier data for q $\|$
[$\xi$00]~\cite{Ti}. The calculations of the phonon self-energies in Ti for bare phonon frequency $\omega_0$ (Table~\ref{tab1}) show that in this case probed momenta
distribution located mainly in the NA region (Fig.~\ref{fig5}). This is the reason of rather strong
($\simeq 8\%$) frequency softening and non-monotonic behavior of width (see temperature dependence of $\Pi''$ in Fig.~\ref{fig5}) when temperature increases. 
Interpolation of high-temperature neutron data~\cite{st1} (Fig.~\ref{fig4}(d)) shows strong deviation from Raman phonon frequencies at low temperatures. Calculations  for different $q$ directions gave almost the same results (solid lines in Fig.~\ref{fig4}(c,d)) which describe experimental phonon frequencies and widths rather good. Nevertheless, found in experiment differences in low -temperature widths suggest contribution of the intraband effects. It is interesting that comparable
quantitative NA effect on the  phonon frequency in Ti was calculated  in Ref.~\cite{mauri}, but the main source of such phonon renormalization was  attributed to interband transitions.

\subsection{Group VI metals - ruthenium and osmium}
\begin{figure}[t] \includegraphics[width=80mm]{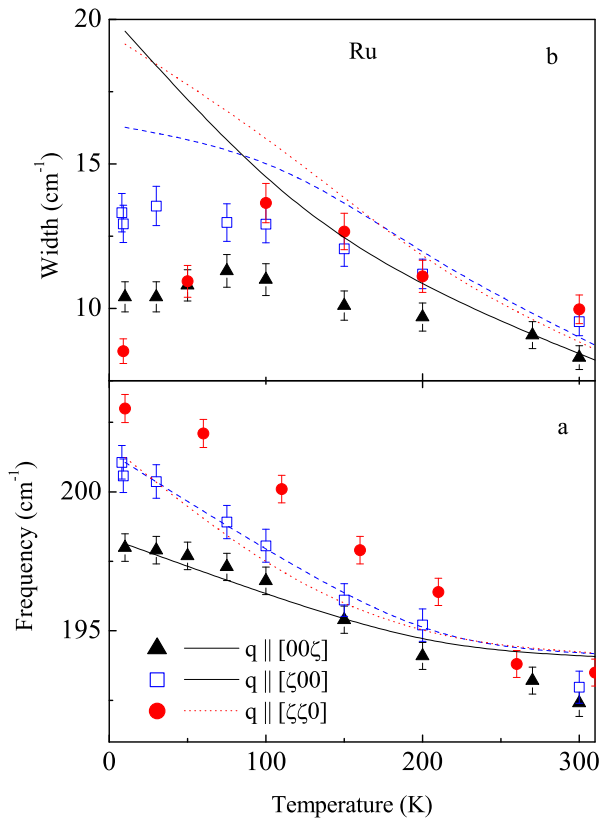}
\caption{\label{fig6} Experimental (symbols) and calculated (lines)
temperature dependence of the phonon frequency (a) and linewidth (b) in ruthenium
for different directions of the wavevector $q$. The excitation wavelength was 647~nm.}
\end{figure}

\begin{figure}[b] \includegraphics[width=90mm]{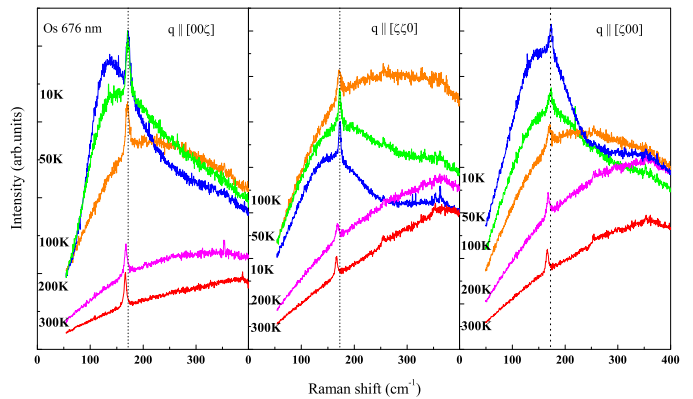}
\caption{\label{fig7}  Raman spectra in osmium measured at different temperatures for different q-directions. Excitation was 676 nm.}
\end{figure}

\begin{figure}[t] \includegraphics[width=90mm]{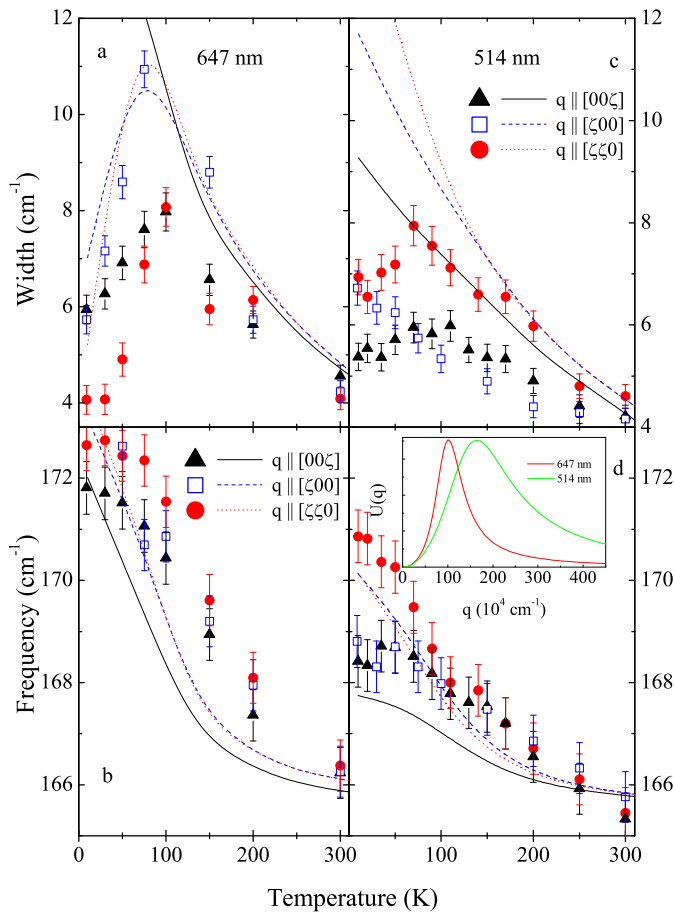}
\caption{\label{fig8} Experimental (symbols) and calculated (lines)
temperature dependence of the phonon frequency and linewidth in osmium for two excitation energies and for different directions of the wavevector. The inset in (d) shows phonon momentum distributions for the two excitation energies.}
\end{figure}

Figures~\ref{fig6}, ~\ref{fig7} and  ~\ref{fig8} show the temperature dependences of the spectra, phonon
frequencies and widths for 4d and 5d metals of VI group Ru and Os. Calculated
electronic structures and Fermi surfaces for both metals are very similar.
Therefore it is not surprising that we found similar temperature behavior of
their phonon self-energies.
In ruthenium NA effects in electron-phonon interaction
are observed through the differences of the phonon self-energies for different
wavevector directions.  We found that at low-temperature the frequencies of
E$_{2g}$  phonon for q $\|$ [00$\xi$] are by 4-5 cm$^{-1}$ less than for in-plane momentum directions, 
but they become almost equal at room temperature (Fig.S4 and Fig.~\ref{fig6}). The frequencies decrease
 with increasing temperature for
all momentum directions, while the widths sometimes show a non-monotonic
dependence, especially for q $\|$[$\xi\xi$0], where the low-temperature
width is close in magnitude to the one observed at 300~K. Calculated  phonon
frequencies and widths, also shown in Fig.~\ref{fig6}, satisfactorily reproduce the anisotropy
in the temperature dependence of frequencies, but show monotonic increase for all q-directions in the contrast experimental
data for q $\|$[$\xi\xi$0]. For this direction both the adiabatic and NA regions contribute to peak intensity that
leads to large growth of the linewidth at low temperatures (Fig.S4 and Fig.~\ref{fig6}).

The variation of osmium refraction index n and extinction coefficient k with
wavelength allows the wavevector q to be tuned in the region (6-18)$\cdot10^5$ cm$^{-1}$. 
Such variation leads to the changes in the
frequencies and widths of the phonon lines for certain directions of
wavevector~\cite{pbtc} (Fig.~\ref{fig7} and ~\ref{fig8}). Particularly noticeable the changes
in low-temperature frequencies for directions of q $\|$[00$\xi$] and q
$\|$[$\xi$00]. They increase during the transition to the longer-wavelength
excitation and a corresponding decrease in the probed wavevector (see inset
in Fig.~\ref{fig8}). Phonon lines in Os Raman spectra are superimposed on broad electronic background~\cite{pbtc}, 
which is rather structureless for 514 nm excitation but show very expressed profile for
 red line excitation at 647 and 676 nm (Fig.~\ref{fig7}). The maximum of this intraband continuum is close 
to the energy of $E_{2g}$ phonon and its temperature behavior is related to the phonon self-energy trend. It is not surprising
because the frequency and temperature dependence of inelastic light scattering by electronic excitations is determined by imaginary part of Eq.3 where the matrix element of electron-phonon interaction ${g}$ is replaced by matrix element of electron-photon interaction~\cite{carip}.
As with other investigated metals, the anomalous
broadening of the phonon lines upon cooling was observed in osmium. A
pronounced maximum at T $\approx$ 100~K was found for
direction q $\|$ [$\xi\xi$0], where the low-temperature linewidth was
close to its room temperature value for 647 nm excitation, as in the
case of ruthenium (Fig.~\ref{fig6}). As one can see, the intensity of electronic continuum for
 this direction also shows nonmonotonic behavior in this temperature range. This similarity confirms that
 intraband electronic excitations determine the phonon self-energy.
The general trend found is an increase in the low-temperature frequencies and widths of the phonon lines with decreasing
wavevector probed. This clearly indicates the growing contribution of NA
region of wavevectors to the measured phonon frequencies and  widths.
We performed calculations of the phonon spectral functions at different
temperatures (Eqs.~\ref{eq2}-\ref{eq6}) for both excitation
wavelengths. The results obtained (Fig.~\ref{fig8}) reproduce
observed at low temperatures phonon softening and line narrowing, when the
probed momentum increases. However, it is not possible to describe a line
narrowing at low temperatures for q $\|$ [00$\xi$] direction upon excitation
at 647 nm. Also the low temperature linewidths for excitation at 514 nm are overestimated due to
arising high energy shoulders in calculated spectral functions. 

\section{Discussion}

Thus, the narrowing of the phonon lines or non-monotonic behavior with increasing temperature is observed for all investigated metals. Typically, the phonon frequencies decrease with increasing temperature. Temperature behavior of the phonon parameters depends on the direction of probed wave vector which. The lowest anisotropy was found in titanium, where the frequencies for q-directions practically coincide. Found q - dependences imply the contribution of the intraband electronic transitions to the observed effects. The proposed explanation of these effects is based on the account of electron spectrum renormalization due to carrier scattering by phonons.

The main contribution to the narrowing of the phonon lines is due to the increase of the electron relaxation
rate with increasing temperature. This leads to a decrease the time during which the electron moves in phase with the phonon and interacts effectively with it in the adiabatic region of the spectrum.  Otherwise, in the NA region the phonon damping increases at intermediate temperatures due to a decrease of the electron lifetime. Interplay of these mechanisms sometimes results in a non-monotonic temperature behavior of the total phonon widths.

Possible reasons for discrepancies between the experimental and calculated results
may be inaccuracies in the calculations of the Fermi surface shape and the
distribution of electron velocities on it. In addition, a rough calculation using the isotropic
renormalization of the electron spectrum and the same matrix elements of
electron-phonon interaction for all Fermi surface sheets, may not produce the subtle
features in the behavior of the phonon self-energies. Nevertheless, overall
decrease of the frequencies and linewidths upon transition to shorter excitation wavelength in Os
was reproduced in calculations.

There are two additional unaccounted effects which could distort the
calculated spectral functions for all metals. One of them is the neglecting
of the influence of screening in the case of longitudinal phonons for the
 momentum direction along [$\xi\xi$0]. The point group $D_{6h}$ of a hcp crystal leads to first-order
 Raman activity for only one doubly degenerate zone-center optical phonon mode of $E_{2g}$ symmetry.
In this vibration, two sublattices undergo opposite displacements in the $x$- and $y$-directions of the
basal plane. For small finite wavevector along the [00$\xi$]
direction (measurements from (0001) surface) the doubly degenerate transverse
phonon is probed, while the longitudinal mode in [$\xi\xi$0] direction and
the transverse mode in [$\xi$00] direction may be studied in measurements from
(1010) and (1210) planes. It is possible that the difference in low-temperature phonon frequencies for different q 
in the basal plane in Os and Ru (see Figs.6-8) is associated with this effect.
The second unaccounted effect is associated with a possible
line shape change of the phonon lines due to the Fano interference with the
electronic continuum. 

In all performed calculations we did not take into
account another possible sources of phonon softening with increasing temperature
which are thermal expansion and higher anharmonic processes. Estimates made
show the thermal expansion effects do not exceed 10\% of the observed phonon softening in all investigated
metals. Some anharmonic contributions to the phonon damping and energy were
observed in Ti at $T \geq$ 300~K; no linewidth broadening was found in Ru and
Os up to 800~K. Therefore, we believe that the main part of the temperature
changes in phonon self-energies are associated with electronic contributions.

\section{Conclusions}

We present new results on the temperature and momentum dependent optical $E_{2g}$
phonon self-energies in five hcp transition metals. The comparison of
experimental and calculated phonon spectral functions clearly evidences that
the nonadiabatic effects in electron-phonon interaction strongly contribute to
detected anomalies. According to our
calculations, the main part of the effects comes from the intraband mechanism.
The contribution of the interband transitions still needs to be estimated.
Despite differences of experimental and calculated phonon parameters,
we were able qualitatively (and sometimes even quantitatively) to describe
the observed temperature behavior of the self-energies of the long wavelength
optical phonons and their anisotropy for all investigated metals.  Though
these effects amount to a few percent, however, they determine the temperature
dependence of the phonon frequencies, which can be mistakenly identified with
the anharmonic behavior. 

As a general remark, the ability of a Raman experiment to probe the nonadiabatic effects on
the optical phonon self-energy requires a careful analysis of
temperature dependence of phonon self-energies in metals and doped
semiconductors to separate the effects of phonon-phonon and electron-phonon
interactions.

\section*{Acknowledgments}

We thank K. Syassen for valuable discussions. The research was supported by
the grant of the Russian Scientific Foundation (project no. 14-22-00004).

\end{document}